\documentclass[reprint,aps,prl,amsmath,amssymb,floatfix]{revtex4-1}

\pdfoutput=1

\usepackage{graphicx}
\usepackage[pdftex,breaklinks=true,bookmarksopen=true,bookmarksopenlevel=3,bookmarksnumbered=true,colorlinks=true,urlcolor= magenta,citecolor=blue,linkcolor=blue]{hyperref}

\begin{document}

\title{Controlled Betatron X-Ray Radiation from Tunable Optically Injected Electrons}

\author{S.~Corde}
\author{K.~Ta Phuoc}
\author{R.~Fitour}
\author{J.~Faure}
\author{A.~Tafzi}
\author{J.~P.~Goddet}
\author{V. Malka}
\author{A.~Rousse}
\affiliation{Laboratoire d'Optique Appliqu\'ee, ENSTA ParisTech - CNRS UMR7639 - \'Ecole Polytechnique, Chemin de la Huni\`ere, 91761 Palaiseau, France}

\begin{abstract}
The features of Betatron x-ray emission produced in a laser-plasma accelerator are closely linked to the properties of the relativistic electrons which are at the origin of the radiation. While in interaction regimes explored previously the source was by nature unstable, following the fluctuations of the electron beam, we demonstrate in this Letter the possibility to generate x-ray Betatron radiation with controlled and reproducible features, allowing fine studies of its properties. To do so, Betatron radiation is produced using monoenergetic electrons with tunable energies from a laser-plasma accelerator with colliding pulse injection [J. Faure \emph{et al.}, \href{http://dx.doi.org/10.1038/nature05393}{Nature (London) \textbf{444}, 737 (2006)}]. The presented study provides evidence of the correlations between electrons and x-rays, and the obtained results open significant perspectives toward the production of a stable and controlled femtosecond Betatron x-ray source in the keV range.
\end{abstract}


\maketitle

The continuous progress made over the past decade in the production of ultrashort x-ray radiation has opened novel research horizons with countless applications. The most advanced short pulse x-ray source to date is the free electron laser (FEL), producing x-ray pulse orders of magnitude brighter than any other source~\cite{NatPhot2010Emma}. However, at such large facilities, beam time access is inevitably limited and there is therefore an interest in producing compact sources delivering x-ray pulses, even if less intense than in a FEL, sufficiently bright to satisfy the need of many applications. For this reason, the research on complementary femtosecond x-ray sources remains dynamic and novel sources are developed in both synchrotron and laser-plasma interaction communities.
Produced in relativistic laser-plasma interaction, the Betatron radiation presents promising features to become a bright and compact femtosecond x-ray source. Demonstrated in 2004, this scheme allowed us to produce for the first time broadband low divergence femtosecond x-ray beams in the keV energy range from laser-plasma interaction~\cite{PRL2004Rousse}. Since then the source has been developed and widely characterized. It can now generate radiation with divergence down to below 10 mrad and in the 10 keV range~\cite{NatPhys2010Kneip}. However, since its first demonstration, the Betatron radiation has always been produced by self-injected electrons in the bubble or blowout regime~\cite{PRL2004Rousse, PRL2006TaPhuoc, PoP2007TaPhuoc, PRE2008Albert, APL2009Mangles, NatPhys2010Kneip}. In that case, with present laser technologies, the laser pulse has to be shrunk in time and space to generate an appropriate bubble structure. These effects, that result from the relativistic self-focusing and self-shortening, are strongly nonlinear and by nature unstable; small fluctuations of any of the experimental parameters will likely lead to important fluctuations of the Betatron radiation properties. Self-injected laser-plasma accelerators do not allow us to easily control the electron and radiation properties. In addition, correlations between the electron beam energy and x-ray properties are difficult to observe because the relevant parameters can not be controlled independently (the increase of electron beam energy usually comes with a decrease of the plasma density and transverse amplitude of motion). A solution to improve the control on the electron and radiation properties is to use the recently demonstrated laser-plasma accelerator in the colliding pulse geometry ~\cite{PRL1996Umstadter, PRL1997Esarey, PRE2004Fubiani, Nature2006Faure, PRL2009Rechatin1}. In this scheme, electrons are optically injected, in a controlled way, into a moderately nonlinear wakefield and the produced electron bunch has a stable and tunable energy. This colliding pulse injection scheme has the advantage of decoupling the injection process from the acceleration, and hence to produce electron beams with controlled energy and small energy spread while keeping the same plasma density, which allows the precise study of the physics of Betatron radiation and to control its properties. 

In this Letter, we demonstrate experimentally that Betatron radiation can be generated with controlled features using electrons produced by colliding pulse injection. In addition, we present the first precise study showing clear correlations between the electron beam and the Betatron radiation. This has allowed us to define a relevant theoretical model to determine the source features and its scaling laws. As a consequence, it opens perspectives for future developments. 

In a laser-plasma accelerator, the Betatron radiation is emitted by electrons which are accelerated up to relativistic energies and undergo transverse oscillations in the wakefield of an intense short pulse laser propagating in an underdense plasma. The features of the Betatron radiation (angular and spectral) uniquely depend on the electron orbits in the accelerator. A relevant parameter to describe the Betatron source is the strength parameter $K=r_\beta k_p\sqrt{\gamma/2}$, with $r_\beta$ the transverse amplitude of the electron orbit, $k_p=\omega_p/c$ with $\omega_p$ the plasma frequency and $\gamma$ the electron Lorentz factor. In the wiggler regime, $K\gg1$ (regime of all experiments performed so far), the radiation consists of a low divergence beam with a broadband synchrotron type spectrum and a pulse duration of the same order as the electron bunch duration. For an electron oscillating with a constant energy $\mathcal{E}=\gamma mc^2$ and constant $K$, the critical energy $E_c$ of the synchrotron spectrum, the half divergence $\theta$ (in the plane of motion) and the number of emitted x-ray photons per Betatron period $N_X$ [at the mean energy $\langle E \rangle=8E_c/(15\sqrt{3})$], are given by $E_c = 3 K  \gamma^2 \hbar\omega_\beta/2$, $\theta = K/\gamma$ and $N_{X} = 3.3 \times 10^{-2}K$, where $\omega_\beta=\omega_p/\sqrt{2\gamma}$ is the Betatron frequency.

\begin{figure}[b!]
\includegraphics[width=8.5cm]{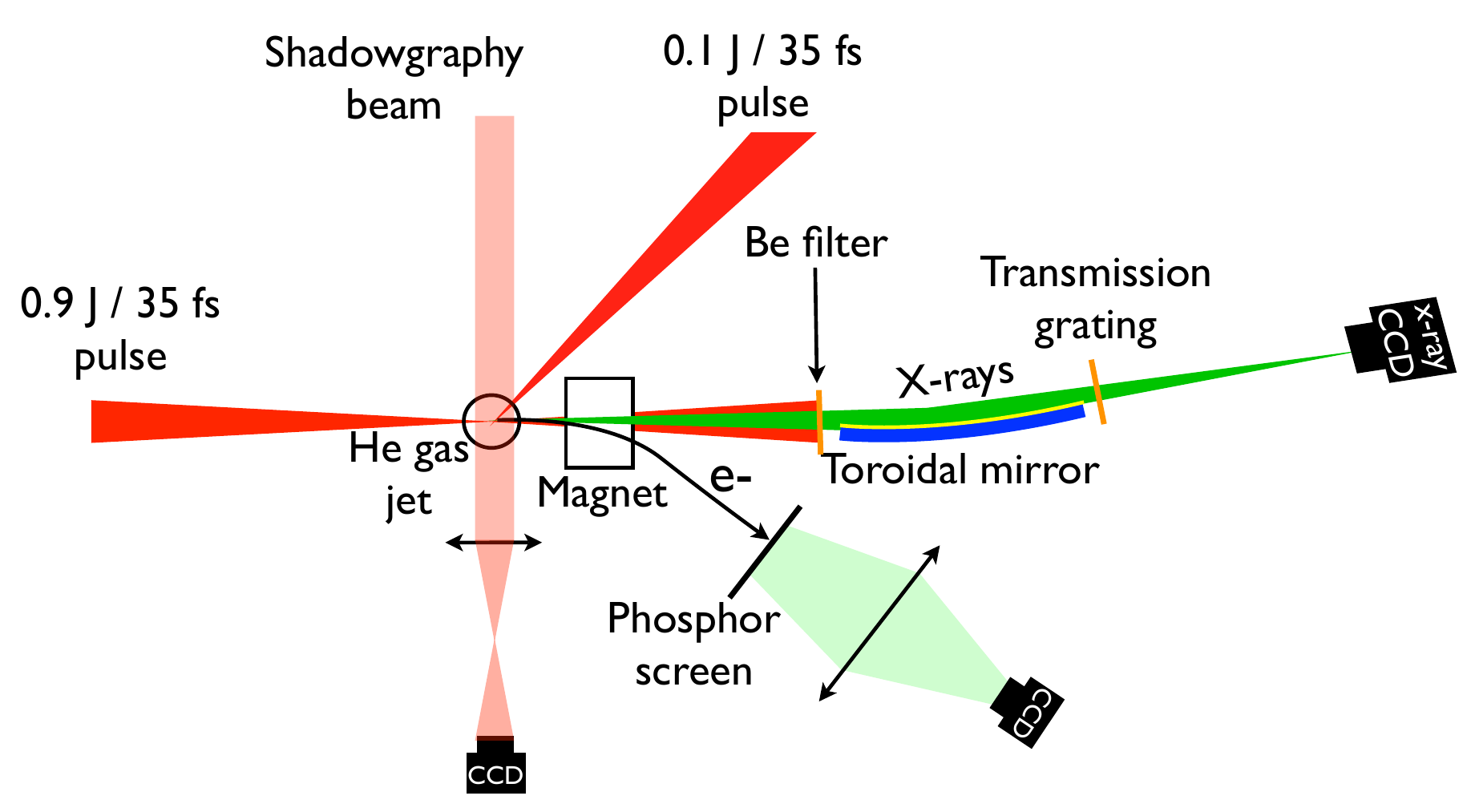}
\caption{Schematic diagram of the experimental setup.}
\label{fig1}
\end{figure}
The experiment was performed at Laboratoire d'Optique Appliqu\'ee with the ``Salle Jaune'' Ti:Sa laser system. Two synchronized 35 fs [full width at half maximum (FWHM)] laser pulses were used: the pump pulse that creates the accelerating plasma wave contained 900 mJ and the injection pulse that triggers the injection into the main pump pulse wakefield contained 100 mJ. The two pulses had the same linear polarization. The experimental setup is presented in Fig. \ref{fig1}. The two beams, making a 135 degrees angle, were focused onto a supersonic helium gas jet (3 mm diameter). For the pump pulse (respectively the injection pulse), the FWHM focal spot size was 18~$\mu$m (respectively 22~$\mu$m) and the normalized vector potential amplitude was $a_0=1.3$ (respectively $a_0=0.4$). Electron spectra and x-ray spectra or angular profiles were measured simultaneously in a single shot. The electron spectrometer consisted of a permanent bending magnet (1.1~T over 10~cm) combined with a phosphor screen imaged on a 16 bits CCD camera.
The x-ray spectrometer consisted of a toroidal mirror imaging the x-ray source at 1~m on the x-ray CCD camera, combined with a 5000~g/mm transmission grating. The x-ray spectrometer acceptance solid angle is limited by the grating size (1~mm~$\times$~1~mm) to $\Omega=1.1\times10^{-5}$~sr. Compared to diagnostics based on filters~\cite{PRL2004Rousse, APL2009Mangles, NatPhys2010Kneip} or crystals~\cite{PRE2008Albert}, this grating spectrometer allows us to record single shot x-ray spectra, with a good resolution and over a large bandwidth, from about 1 to 4~keV, a lower photon energy range than what was recently measured by the photon counting method~\cite{NJP2011Fourmaux}. For x-ray angular profile measurements (not shown in Fig. \ref{fig1}), the spectrometer was replaced by an x-ray CCD placed on axis at 90~cm from the gas jet. The laser light was blocked with a 25~$\mu$m Be filter. In this experiment, the electron plasma density was $n_e = 8\times10^{18}$ cm$^{-3}$, which corresponds, for our experimental parameters, to an interaction regime where electrons are not self-injected in the wakefields. Consequently, electrons and x-rays were observed only when both laser pulses overlapped in time and space.

We controlled the acceleration length $L_\text{acc}$ and thus the final electron beam energy $\mathcal{E}$ by changing the collision position $z_\text{col}$ between both laser pulses ($z_\text{col}=0$ corresponding to a collision at the gas jet center). Doing so, it was possible to tune the electron bunch energy from below 100~MeV to above 200~MeV. The electron beam divergence and charge were typically on the order of 5~mrad (FWHM) and 20~pC. The features (divergence, flux and spectrum) of the Betatron radiation were studied as a function of the electron energy.

\begin{figure}[b!]
\includegraphics[width=8.5cm]{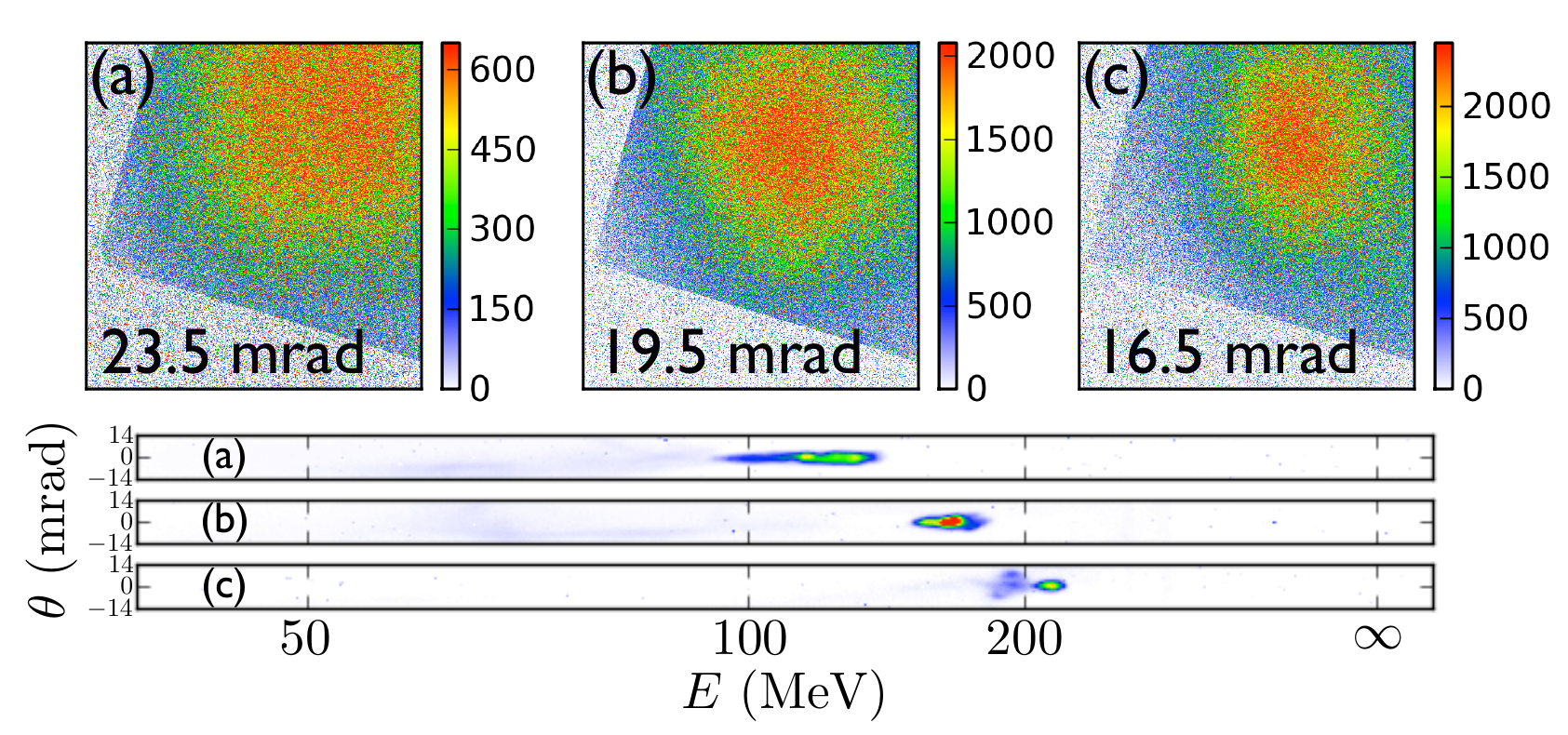}
\caption{Sample of simultaneously measured x-ray angular profiles [top (a), (b), (c)] and raw electron spectra [bottom (a), (b), (c)], for different collision positions $z_\text{col}$. The field of view in each x-ray image is 29.8~mrad~$\times$~28.9~mrad and the color scale gives the number of counts recorded by the x-ray CCD. The geometric average of the horizontal and vertical FWHM x-ray beam divergences is indicated on each x-ray image. In raw electron spectra, the horizontal axis gives the electron energy, the vertical axis the exit angle and the color scale the number of counts (giving an indication of the beam charge).}
\label{fig2}
\end{figure}
The first x-ray property which is found to be correlated with the final electron beam energy is the x-ray beam divergence.
Figure \ref{fig2} presents electron spectra and x-ray angular profiles simultaneously recorded, for different collision positions $z_\text{col}$. The divergence of the measured x-ray angular profiles has a clear dependance on the final electron beam energy: it decreases as the electron energy increases. For electron beams with peak energies at 113, 161, and 221~MeV, the measured x-ray beam FWHM divergences are respectively $\theta = 23.5$, $\theta = 19.5$, and $\theta = 16.5$~mrad. Note that a change in the electron beam transverse size can also have an effect on the observed behavior.

 \begin{figure}[b!]
\includegraphics[width=8.5cm]{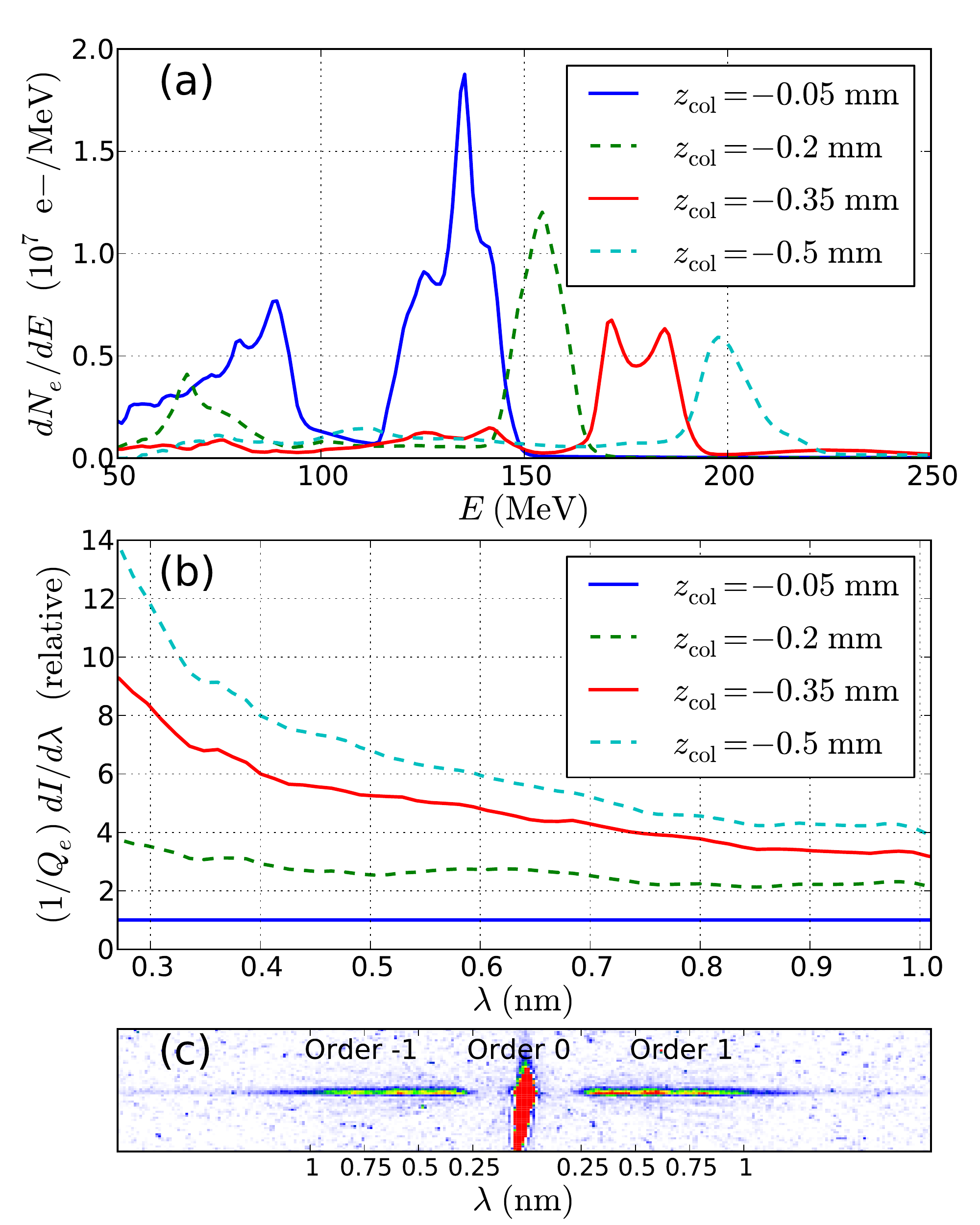}
\caption{Sample of simultaneously measured electron (a) and x-ray (b) spectra, for different collision positions $z_\text{col}$. x-ray spectra are given per unit electron beam charge and are normalized by the first x-ray spectrum obtained at $z_\text{col}=-0.05$~mm (blue curve). The relative rms error on the relative spectra is 5\% at 0.4 nm. (c): A raw x-ray CCD image showing the grating diffraction orders, from which x-ray spectra are obtained.}
\label{fig3}
\end{figure}
The x-ray photon energies and the radiated energy per solid angle strongly depends on the electron energy. Figure \ref{fig3} shows four electron spectra and the corresponding Betatron x-ray radiation spectra, for different collision positions $z_\text{col}$. As x-ray photons were collected on axis in the small solid angle $\Omega=1.1\times 10^{-5}$~sr, the recorded spectrum is approximately the on axis spectrum, \textit{i.e.} $\simeq d^2I/(d\lambda d\Omega)_{|\theta=0}\: \Omega$. To present the evolution of the Betatron radiation spectrum as a function of the final electron beam energy, we represent relative spectra, \textit{i.e.} spectra divided by a reference spectrum (chosen here as the low electron beam energy x-ray spectrum, blue curve in Fig. \ref{fig3}). This permits to remove systematic errors associated to the x-ray spectrometer calibration. In addition, x-ray spectra are divided by the electron beam charge. The bottom part of the figure presents a raw image recorded by the x-ray CCD camera in a single shot. It shows the different orders of the diffraction grating and confirms that the Betatron radiation is emitted in the wiggler regime: x-ray spectra are broadband and extend in the few keV range.

\begin{figure}[b!]
\includegraphics[width=8.5cm]{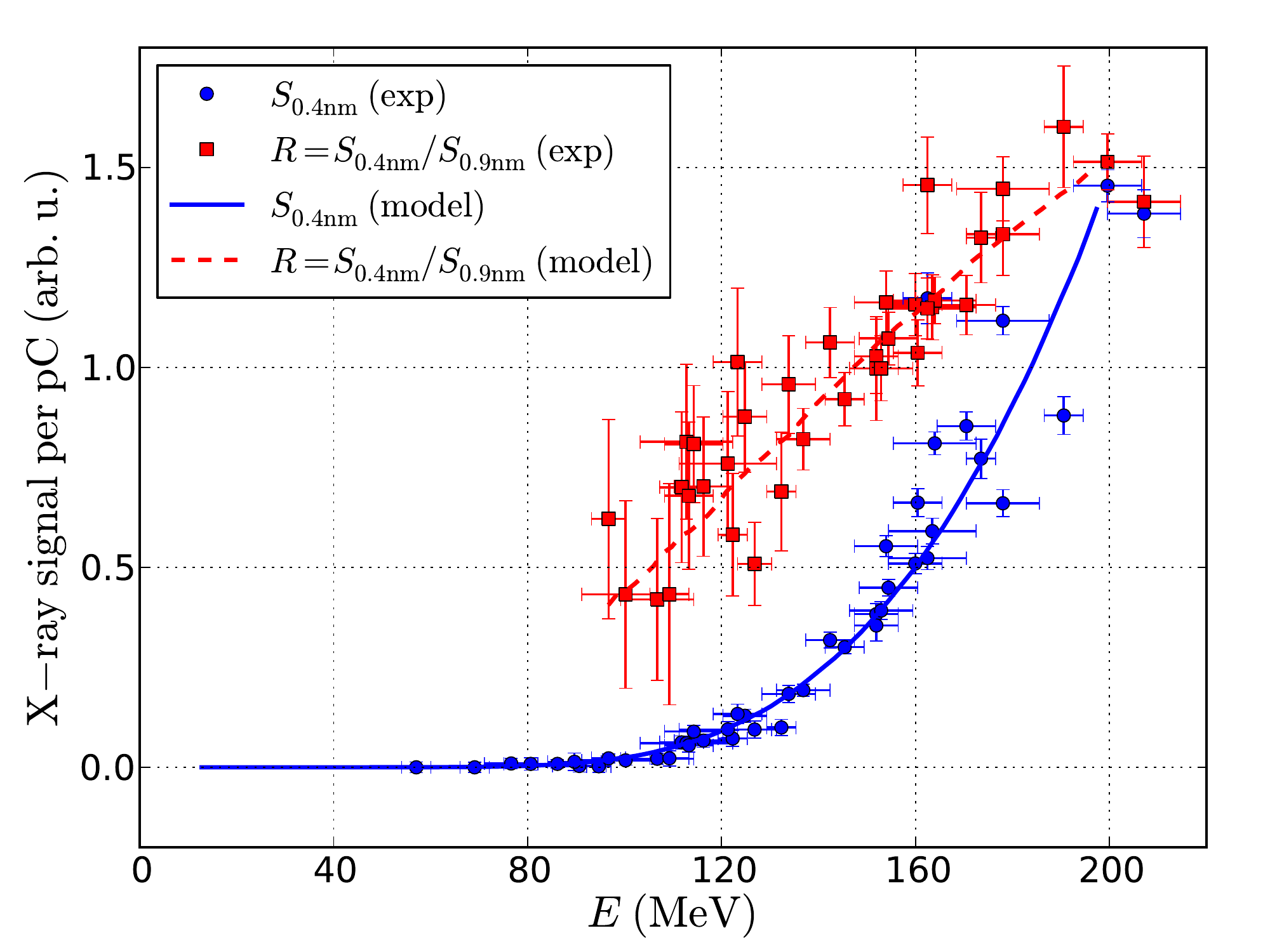}
\caption{Evolution of the x-ray signal at 0.4~nm per unit electron beam charge, $S_{0.4\text{nm}}\propto (1/Q_e)dI/d\lambda_{|0.4\text{nm}}$ (blue circle), and the ratio $R=S_{0.4\text{nm}}/S_{0.9\text{nm}}$ between the x-ray signal at 0.4 and 0.9~nm (red square), as a function of the electron beam energy. Horizontal error bar extremities are positioned at half maximum of electron spectra, and vertical error bars show the estimated error on the x-ray signal measurement. The results of the model are shown in blue solid line and red dashed line, respectively for $S_{0.4\text{nm}}$ and $R$.}
\label{fig4}
\end{figure}
The x-ray spectrum exhibits a strong dependence with the electron beam energy. We define the experimental observable $S_{\lambda_0} = (c_{\lambda_0}/Q_e)dI/d \lambda_{|\lambda_0}$ as the x-ray signal measured at $\lambda=\lambda_0$ divided by the electron beam charge $Q_e$ in the main peak (neglecting the effect of low energy electrons present in some electron spectra), where $c_{\lambda_0}$ is a calibration factor and $dI/d \lambda$ the absolute x-ray spectrum (in J/nm). First, at a given wavelength $\lambda_0$, the x-ray signal per unit electron beam charge $S_{\lambda_0}$ increases rapidly with the final electron energy. Second, the signal increases much more at short wavelength, than at long wavelength. This corresponds to a change in the spectrum shape, which shifts towards shorter wavelengths (higher photon energies) when the final electron beam energy increases. 

For a more quantitative study of the evolution of the x-ray spectral properties with the final electron beam energy, the x-ray signal $S_{0.4\text{nm}}$ and the ratio  $R=S_{0.4\text{nm}}/S_{0.9\text{nm}}$ are plotted in Fig. \ref{fig4} as a function of the electron beam energy for a full set of shots with different collision positions $z_\text{col}$. The x-ray signal $S_{0.4\text{nm}}$ provides information on the evolution of the radiated energy per unit solid angle at a fixed wavelength, while the ratio $R$ is representative of the shift of the spectrum towards lower or higher photon energies. The result shows again that these quantities are strongly correlated with the electron beam parameters. The x-ray signal $S_{0.4\text{nm}}$ is smaller than the detection threshold for electrons below 80~MeV, and then increases sharply with the electron beam energy. In addition, the increase of $R$ with the electron energy shows how the x-ray spectrum shifts towards higher photon energies when electron energies are increased. The sharp increase of $S_{0.4\text{nm}}$, as observed in Fig. \ref{fig4}, justifies \textit{a posteriori} that we can neglect the influence of low energy electrons present in some electron spectra (e.g. see the spectrum at $z_\text{col}=-0.05$ mm in Fig. \ref{fig3}), because they lead to a contribution to the x-ray signal smaller than the error bar of the measurement.

The behaviors of $S_{0.4\text{nm}}$ and $R$ can be explained as follows. The measured value of $S_{\lambda_0}$ can be written as a function of the final Lorentz factor $\gamma_f$ of the electron beam as
\begin{align}
S_{\lambda_0}(\gamma_f) &= \frac{c_{\lambda_0}}{Q_e} \int_1^{\gamma_f}\frac{hc}{\lambda_0^2\dot{\gamma}}\frac{dP}{dE}_{|E_0}(\gamma)\,d\gamma,
\label{eq1}
\end{align}
where $\dot{\gamma}=d\gamma/dt$ and $dP/dE_{|E_0}(\gamma)$ is the energy radiated per unit time in the photon energy band $dE$ centered on $E_0=hc/\lambda_0$ and in the solid angle $\Omega$, by an electron beam of charge $Q_e$ during its acceleration. Since $dP/dE_{|E_0}$ integrates the contribution of all electrons (having different Betatron amplitudes), it does not strictly have a synchrotron shape but for the sake of simplicity and for the following qualitative discussion we will use this assumption. As the electron beam accelerates and $\gamma$ increases, $S_{\lambda_0}$, with $\lambda_0$ in the subnanometer range, increases for three reasons. First, the increase of the final electron energy comes from an increase of the acceleration length such that $\gamma_f$ appears to be the upper limit of the integral in Eq. (\ref{eq1}), resulting in a necessarily increasing x-ray signal and in a linear increase for a constant integrand. Second, higher electron energies result in a higher critical energy $E_c$ for the photon synchrotron spectrum appearing in the integrand of Eq. (\ref{eq1}), due to the Doppler shift factor of $2\gamma^2$. At the beginning (low $\gamma$), the photon energy $E_0=hc/\lambda_0$ lies in the tail of the synchrotron spectrum, which has an exponential decay $\sim \exp(-E/E_c)$ for $E \gg E_c$. Thus, when $\gamma$ and $E_c$ increase, the integrand increases roughly as $\exp(-E_0/E_c)$ and eventually deviates from this behavior as $E_0$ becomes closer to the critical energy and approaches the peak of the synchrotron spectrum. The experimental behavior of $R$ is the result of the Doppler shift and the increase of $E_c$. Third, the x-ray beam divergence decreases with $\gamma$ and therefore the radiated energy is confined in a smaller solid angle.

These effects can be properly and quantitatively taken into account using a simple model which describes the radiation from an accelerating electron bunch in a focusing field and fits its properties to the experimental results. 
The wake of the laser pulse is approximated by a spherical ion cavity propagating in the $\vec{e}_z$ direction with a velocity close to the speed of light. The motion of electrons trapped in this wake consists of an acceleration in the longitudinal direction $\vec{e}_z$, due to a longitudinal force $\vec{F}_z$, combined with a transverse oscillation across the cavity axis at the Betatron frequency $\omega_{\beta} \simeq \sqrt{\alpha\omega_p^2/(2\gamma)}$~\cite{PRE2002Esarey, PoP2003Kostyukov}, due to a transverse linear focusing force $\vec{F}_\bot=-m\alpha\omega_p^2r/2\: \vec{e}_r$ ($\alpha$ is an adjustable numerical factor used to describe any deviation from the nominal transverse force corresponding to a fully evacuated ion cavity). Because the electron motion is essentially longitudinal, $p_\bot \ll p_z$, the Hamiltonian describing the electron dynamics can be expanded and separated into a longitudinal and a transverse part~\cite{PoP2004Kostyukov}, $\mathcal{H}=\mathcal{H}_z+\sum_{a=x,y}\mathcal{H}_a$, with
\begin{align}
\mathcal{H}_a &= \frac{p_a^2}{2\gamma m}+\frac{1}{4}m\alpha\omega_p^2a^2=J_a\omega_\beta,
\end{align}
with $a=x,y$ and where $J_x, J_y$ are the action variables which are conserved for an adiabatic variation of $\gamma$ such that $| (1/\omega_\beta^2)d\omega_\beta/dt |  \ll 1$ (\textit{i.e.} an adiabatic acceleration).

We consider an electron bunch initially described by a Maxwell-Boltzmann transverse distribution function $f(x,y,p_x,p_y)=f_0\exp(-\sum_{a=x,y}\mathcal{H}_a/k_BT_\bot)$ and we numerically solve for the electron trajectories and then compute their radiation from the Li\'enard-Wiechert fields~\cite{Jackson}. In the expression for $f$, $T_\bot$ is the electron beam transverse temperature which evolves as $T_\bot \propto \gamma^{-1/2}$ for an adiabatic acceleration. The longitudinal force is estimated from the measurement of $\mathcal{E}(z_\text{col})$, and is approximately constant, $F_z \simeq 100\: \text{GeV.m}^{-1}$. The transverse initial positions and momenta of $10^3$ electrons are randomly generated according to the distribution function $f$. The model reproduces the experimental behaviors of $S_{0.4\text{nm}}$ and $R$ simultaneously for $\alpha=1$ and  for a rms transverse size $\sigma=\sqrt{2k_BT_\bot/(m\alpha\omega_p^2)}=0.23 \pm 0.08 \: \mu$m at $\mathcal{E}=200$~MeV, as can be seen in Fig. \ref{fig4} [and also for other choices of ($\alpha$,$\sigma$) satisfying $\alpha\sigma=0.23\: \mu$m]. The fit parameters used here are $\alpha\sigma$, $c_{0.4\text{nm}}$ and $c_{0.4\text{nm}}/c_{0.9\text{nm}}$. This result shows that the measured x-rays follow the properties of synchrotron radiation from electrons in acceleration which are oscillating in a transverse focusing force, and provides a measurement of $\alpha\sigma$. This measurement, together with the measurement of other properties of Betatron radiation (beam divergence and source size), can provide a complete characterization of laser-plasma accelerator properties: emittance, focusing strength and width of the angular momentum distribution. This characterization will be the object of a separate paper.

In conclusion, we presented experimental results on the production of Betatron radiation by controlled and stable electron beams generated in a laser-plasma accelerator with optical injection. Spectra or angular profiles were measured in a single shot, simultaneously with the electron beam. The decoupling between the electron injection process and the acceleration offers a unique possibility to explore the physical mechanisms at the origin of the source. Indeed, for the first time, the Betatron radiation is produced by well defined and tunable electron beams, without changing the plasma density, which permits to test the links between electrons and x-rays and to confirm the scaling laws for future developments of the source. The results are well reproduced with a simple model of radiation from an accelerating electron bunch in an ion cavity. They show a strong correlation between the electron beam parameters and the x-ray properties, validating the ion cavity model and the synchrotron nature of the observed radiation. They also give an insight into the accelerator properties, through the measurement of $\alpha\sigma$.

The authors acknowledge the support of the European Research Council (PARIS ERC, Contract No. 226424) and EuCARD/ANAC, EC FP7 (Contract No. 227579).



\end{document}